\documentclass[12pt]{article}
\usepackage{epsfig,cite}
\usepackage{latexsym}
\usepackage{threeparttable}
\usepackage{amssymb,amsmath}

\newcommand{\beq}{\begin{eqnarray}}
\newcommand{\eeq}{\end{eqnarray}}
\newcommand{\be}{\begin{eqnarray*}}
\newcommand{\ee}{\end{eqnarray*}}

\begin{document}

\title{{\large  \hfill CERN-PH-TH/2010-100}\\\vskip 1cm
{\bf Nuclear effects on the longitudinal structure function at small $x$}}
\author{N\'estor~Armesto$^a$, Hannu Paukkunen$^a$, Carlos A.~Salgado$^{a,b}$\\ and Konrad~Tywoniuk$^a$\\
\\
{\it $^a$ Departamento de F\'{\i}sica de Part\'{\i}culas and IGFAE}\\
{\it Universidade de Santiago de Compostela}\\
{\it 15706 Santiago de Compostela, Galicia, Spain}\\
\\
{\it $^b$ Physics Department, Theory Unit, CERN}\\
{\it CH-1211 Gen\`eve 23, Switzerland}
\\
\\
E-mails: {\tt nestor.armesto@usc.es, hannu.paukkunen@usc.es,}\\
{\tt carlos.salgado@usc.es, konrad.tywoniuk@usc.es}}
\date{September 8th 2010}
\maketitle

\begin{abstract}
We discuss the longitudinal structure function in nuclear DIS at small $x$. We work within the framework of 
universal parton densities obtained in DGLAP analyses at NLO.
We show that the nuclear effects on the longitudinal structure function closely follow those on the gluon distribution.
The error analyses available from newest sets of nuclear PDFs also allow to propagate the 
uncertainties from present data.
In this way, we evaluate the minimal sensitivity required in future experiments for this observable to 
improve the knowledge of the nuclear glue. 
We further discuss the uncertainties on the extraction of $F_2$ off nuclear targets, introduced by the usual assumption that the ratio $F_L/F_2$ is independent of the nuclear size. We focus on the kinematical regions relevant for future lepton-ion colliders.
\end{abstract}

\section{Introduction}
\label{intro}

Nuclear effects on the structure functions measured in deep inelastic scattering (DIS) experiments \cite{Arneodo:1992wf,Geesaman:1995yd} offer valuable information for understanding the dynamics of partons in the nuclear environment. At small values of the Bjorken variable $x$, such measurements provide a clean experimental setup for studying the behavior of QCD at high energies (see the review \cite{Armesto:2006ph} and references therein). 

A usual framework for the determination of nuclear parton distribution functions (nPDFs) is that of global analysis using the Dokshitzer-Gribov-Lipatov-Altarelli-Parisi (DGLAP) evolution equations \cite{Dokshitzer:1977sg,Gribov:1972ri,Altarelli:1977zs}. 
The procedure, pioneered in \cite{Frankfurt:1990xz,Eskola:1992zb,Eskola:1998iy,Eskola:1998df}, is 
similar to the ones performed for the proton case: obtaining the parameters for a set of parton distribution functions (PDFs) at an initial scale $Q_0^2$ which best reproduce some given sets of experimental data. The quality criterion is provided by a proper definition of a $\chi^2$-function. As in the proton case, the goal of these analyses is twofold: on the one hand to check the degree of compatibility of different data sets within an approach of universal PDFs evolved by DGLAP evolution equations; and, on the other hand, to provide a tool to compute cross sections for other processes in terms of a released set of PDFs for the different parton species. Given the fact that the factorization theorems in QCD \cite{Collins:1989gx,Qiu:2003cg} are expected to be broken more easily for nuclei than for free protons, these checks are clearly of great importance in the phenomenological analyses of nuclear collider data. The latest sets of nuclear PDFs (nPDFs) obtained by these global fits are available at next-to-leading order (NLO) accuracy \cite{deFlorian:2003qf,Hirai:2007sx,Eskola:2009uj,Schienbein:2009kk}, with state-of-the art error analysis using the Hessian method \cite{Hirai:2007sx,Eskola:2009uj}. All these sets use data on nuclear DIS with charged leptons and Drell-Yan in proton-nucleus collisions. Data on inclusive pion production at high-$p_t$ has also been included in \cite{Eskola:2009uj} providing extra constrains for the gluons without introducing tension among different data sets. Furthermore, this 
compatibility of data within a universal set of nPDFs has also been checked in neutrino DIS on nuclei \cite{Paukkunen:2010hb}\footnote{See, however, Ref. \cite{Schienbein:2007fs} for contradictory 
results.}. It is important to emphasize, however, that the role of the gluon distributions in all these checks is rather marginal (this can be seen in the corresponding error bars for gluons computed in \cite{Hirai:2007sx,Eskola:2009uj}) and further checks of this universality would be most welcome.

The main caveat of these analyses is that an initial condition for $Q^2$-evolution -- not motivated from QCD but parametrized in a form flexible enough to reproduce the available experimental data -- is required (for a different approach, see \cite{Guzey:2009jr,Tywoniuk:2007xy,Armesto:2010kr} and references therein). In this situation, the predictive power of the corresponding PDFs is reliable only in the 
region of $x$ covered by experimental data: the extrapolations of both the central value and even the uncertainty bands outside this region remain linked to the functional form of the initial condition used in the analysis. 
For this reason, the results are not reliable for those parton flavors probed well outside the values of $x$ constrained by the fitted data. 
This is particularly severe in the small-$x$ domain and especially for gluons \cite{Armesto:2006ph,Hirai:2007sx,Eskola:2009uj} which, on the other hand, dominate the cross sections at high energies.

In DIS on nucleon targets, it is well known that an additional constrain on the gluon distribution, on top of QCD scaling violations, comes from measuring the longitudinal structure function $F_L$. This quantity has recently been extracted at HERA \cite{:2008tx,Chekanov:2009na} and the resulting impact on constraining the small-$x$ evolution within the global DGLAP fits is currently under discussion \cite{Thorne:2008aj}. On the other hand, due to the poor determination of the nuclear gluon distribution, the measurements of $F_L$ on nuclear targets would be of great importance both for constraining the glue and for studying the nuclear dynamics at small $x$ \cite{Frankfurt:2005mc}. The existent experimental data 
on $F_L$
are sparse and limited to a reduced kinematical region (see \cite{Ackerstaff:1999ac} and references therein). Actually, studies within perturbative QCD \cite{Cazaroto:2008qh} and model calculations \cite{Armesto:2010kr} show that the corresponding nuclear effects closely follow those on the glue at small $x$. 

Furthermore, a knowledge of $F_L$ is required in order to extract $F_2$ from the measurements of the DIS cross section. In the nucleon case, the most recent combined HERA data \cite{:2009wt} provide the cross section in the region $y>0.6$ where the  limited knowledge about $F_L$ introduces a large uncertainty in the extraction of $F_2$. In nuclear DIS, one usually assumes that the nuclear effects on both $F_2$ and $F_L$ are the same, i.e. that the ratio $F_L/F_2$ is independent of the nuclear size and taken to be the same as that in the nucleon \cite{Whitlow:1990gk}.

The purpose of this note is to analyze, within DGLAP approaches at NLO, the predicted nuclear effects on the longitudinal structure function and the uncertainties introduced in the extraction of $F_2$ from the DIS cross section by our {\it a priori} lack of knowledge of the nuclear effects on $F_L/F_2$. This will be done in Sections \ref{flg} and \ref{uncer} respectively. We will focus on the kinematical regions relevant for future lepton-ion colliders \cite{eic,lhec}. We end with some conclusions.

While our work aims at setting  the uncertainties which would be acceptable by a global DGLAP fit, let us note that at small $x$ and small-to-moderate $Q^2$, fixed-order perturbative QCD which we employ here may not be suitable to describe the partonic structure of hadrons and nuclei, see \cite{Armesto:2006ph} and references therein. Non-linear evolution equations which imply a saturation of partonic densities, offer another alternative. Nevertheless, the study of their impact on the longitudinal structure function has been done only for the proton \cite{Albacete:2009fh}, with the studies for nuclear targets performed within saturation-inspired models \cite{Armesto:2002ny,Kugeratski:2005gx,Cazaroto:2008iy}. We will make some qualitative comments on the expected results of saturation on the uncertainties introduced in the extraction of $F_2$ from the DIS cross section at the end of Section \ref{uncer}.

\section{Nuclear effects on $F_L$}
\label{flg}

We define the nuclear ratio of function $f$ (where $f=F_2,F_L,g,\dots$) for nucleus of mass number $A$ 
at momentum fraction $x$ and squared virtuality $Q^2$ as usually:
\beq
R_f^A(x,Q^2)=\frac{f^A(x,Q^2)}{A\times f^p(x,Q^2)}\,,
\label{nsf}
\eeq
where $p$ stands for proton. At small $x$, experimental data indicate that $R_{F_2}<1$, commonly 
referred to as nuclear shadowing.

In what follows, we will work at NLO in the zero mass $\overline{\mbox{MS}}$ scheme, using CTEQ6.1M PDFs in the proton\footnote{For HKN07, the MRST98 set of free proton PDFs are being used as their code gives absolute nPDFs with this set as a baseline.} \cite{Stump:2003yu} and the corresponding nuclear ratios for the different nPDFs given in \cite{deFlorian:2003qf,Hirai:2007sx,Eskola:2009uj,Guzey:2009jr}. With these sets of PDFs we then compute  the corresponding values of the structure functions $F_2$ and $F_L$ (see the expressions in e.g. \cite{Brock:1993sz}) both in the nucleon and nuclear cases.

\begin{figure}[!htb]
\begin{center}
\includegraphics[width=1.0\textwidth]{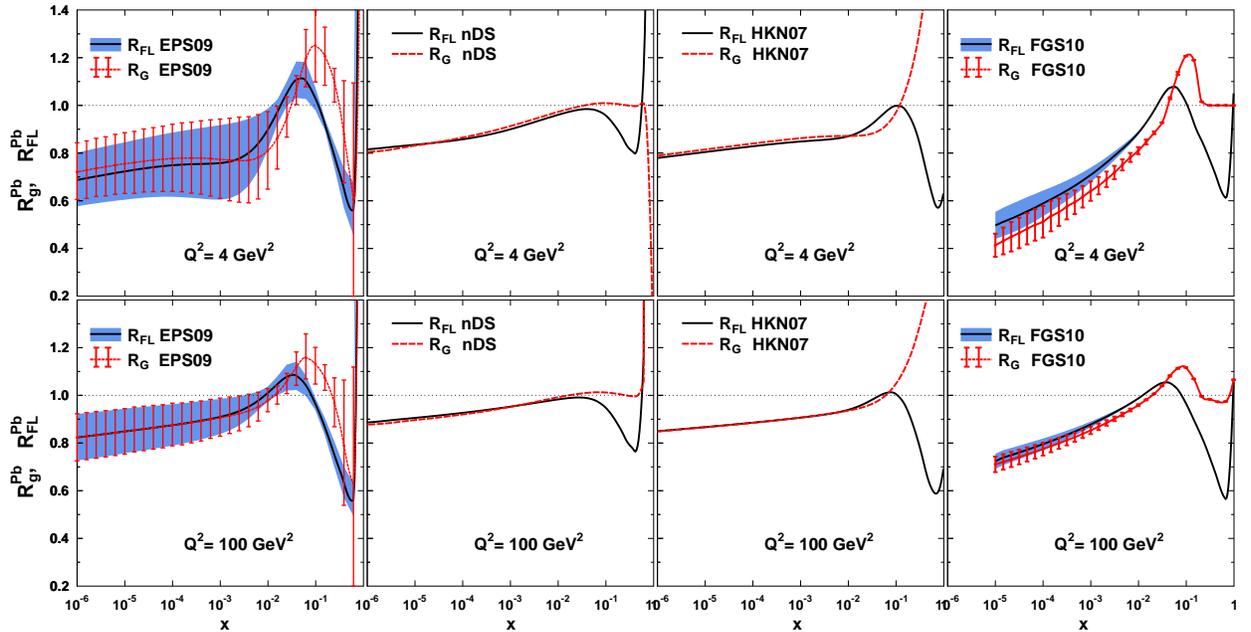}
\caption{Results for $F_L(\mbox{Pb})\big/ F_L$ and for the gluon density in EPS09 (left), and HKN07, nDS and FGS10 (right), for $Q^2=4$ (top) and 100 (bottom) GeV$^2$.
The uncertainty bands and bars in EPS09 correspond to the propagation of the errors from the used data using a Hessian method in the global fit --- see Ref. \cite{Eskola:2009uj} for details. The limits of the bands and bars for FGS10 correspond to two different model implementation in the calculation of  the initial condition and the solid line is an average of these two values. }
\label{fig:flg}
\end{center}
\end{figure}

In pQCD at NLO in the $\overline{\mbox{MS}}$ scheme, the longitudinal structure function $F_L$ has the neat expression
\begin{equation}
 F_L (x,Q^2) = \frac{\alpha_s(Q^2)}{2\pi} \sum_{k=\{q,\overline q \}} e_k^2 \int_x^1 dz \left[ \frac{4}{3} f_k\left(\frac{x}{z},Q^2\right)
+ f_g\left(\frac{x}{z},Q^2\right) (1-z) \right],
\end{equation}
where $f_g$ denotes the gluon PDF, $f_k$'s the corresponding quark PDFs, and $e_k$ is the charge of 
a quark of flavor $k$. With gluon PDFs dominating at small $x$, the nuclear effects on $F_L$ should 
follow those of the gluons. This is clearly seen in Fig. \ref{fig:flg}, where $R^A_{F_L}$ and $R^A_g$ are 
plotted for Pb ($A=208$).
These results agree with the ones in \cite{Cazaroto:2008qh} for those sets used in both analysis. 
Thus a measurement of $F_L$ on nuclear targets offers the possibility of quantifying the nuclear effects on the gluon distribution at small $x$, which are essentially unconstrained in present-day analyses.

\section{Uncertainties in extracting $F_2$ of the nucleus}
\label{uncer}

For neutral current DIS, the measurements are usually given in terms of the reduced cross section, 
which is a combination of structure functions. In the single photon exchange approximation and neglecting, at low and moderate $Q^2$, the contribution 
from electroweak boson interaction, it reads
\beq
\sigma_r^{NC}=\frac{Q^4 x}{2\pi \alpha^2 Y_+} \frac{{d^2} \sigma^{{NC}}}{ {d}x{d} Q^2}=
{F_2}\left[1 -\frac{y^2}{Y_+} \frac{F_L}{F_2}\right],
\label{rcs}
\eeq
where  $\alpha$ is the electromagnetic coupling constant and $Y_{+}=1 + (1-y)^2$.

Thus the extraction of the structure function $F_2$ from $\sigma^{NC}_r$ demands some knowledge on the ratio $F_L/F_2$. The uncertainties introduced by an inaccurate knowledge of this ratio become significant for large $y$ (e.g. for $y>0.6$, $y^2/Y_+>0.31$). With the experimental information about $F_L$ in the proton currently being rather limited \cite{:2008tx,Chekanov:2009na}, the most recent combined HERA data \cite{:2009wt} provide the cross section, and not $F_2$, in the region $y>0.6$.

In the nuclear case, the situation is more complicated by the additional nuclear effects on $F_L/F_2$. The information about the nuclear $F_L$ is sparse and limited at small $0.013<x<0.03$ to $Q^2<1.25$ GeV$^2$, moreover solely for rather light nuclei (see \cite{Ackerstaff:1999ac} and references therein). Previously, in the extraction of $F_2$ from nuclear targets, the usual assumption has been that $F_L/F_2$ is independent of the nuclear size and equal to that in the nucleon\footnote{E. g. using a parametrization \cite{Whitlow:1990gk} extracted from data on the proton and deuteron for $0.1\le x\le 0.9$ and $0.6\le Q^2 \le 20$ GeV$^2$.}.

But it turns out that the recent DGLAP analyses \cite{deFlorian:2003qf,Hirai:2007sx,Eskola:2009uj,Schienbein:2009kk,Guzey:2009jr} of nuclear parton densities give rise to nontrivial effects on the ratio $F_L/F_2$ which, if neglected, may induce additional uncertainties on the extraction of the nuclear $F_2$ from the reduced cross section. In order to estimate this uncertainty, we define the relative uncertainty
\beq
\Delta F_2^{A} = \frac{\tilde F_2^A - F_2^A}{\tilde F_2^A} =1-\frac{\Delta^p}{\Delta^A} \,,
\label{relun}
\eeq
with
\beq
\Delta^{p,A}=1 -\frac{y^2}{Y_+} \frac{F^{p,A}_L}{F^{p,A}_2} \,,
\label{rrr}
\eeq
where $\tilde F^A_2$ is the nuclear structure function extracted under the assumption of no nuclear effects on $F_L/F_2$, while $F_2^A$ is defined by Eq.~(\ref{rcs}).

\begin{figure}[!htb]
\begin{center}
\includegraphics[width=1.0\textwidth]{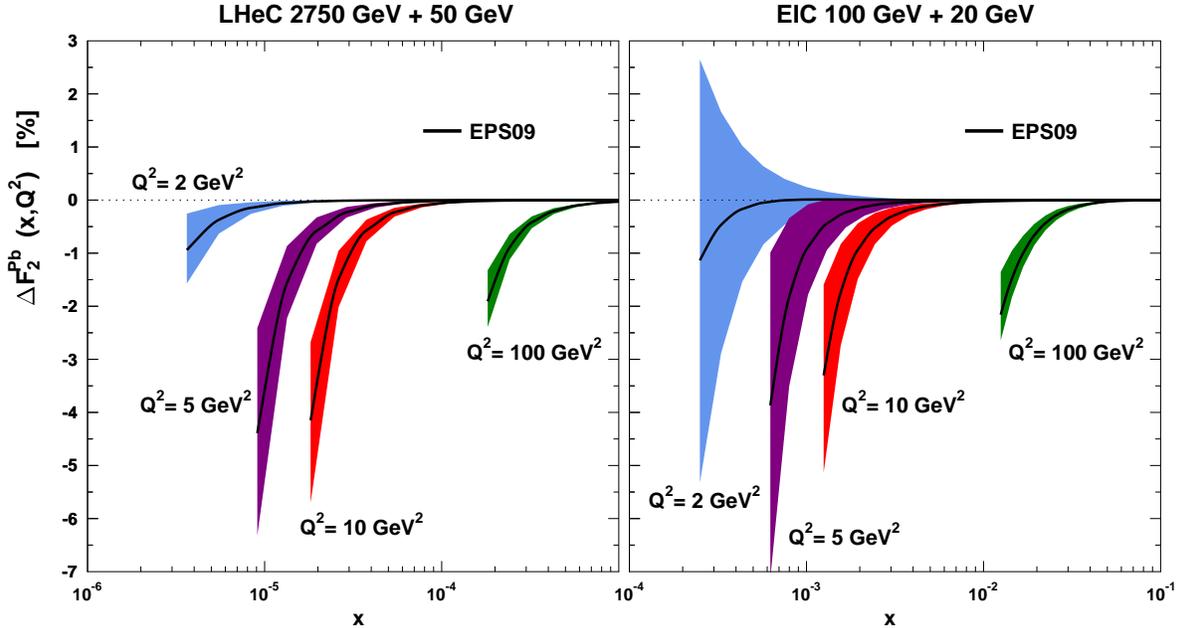}
\caption{Results for the uncertainty in the extraction of $F_2$ in EPS09 for LHeC (left) and EIC (right) kinematics.}
\label{fig:uncertainties}
\end{center}
\end{figure}

\begin{figure}[!htb]
\begin{center}
\includegraphics[width=1.0\textwidth]{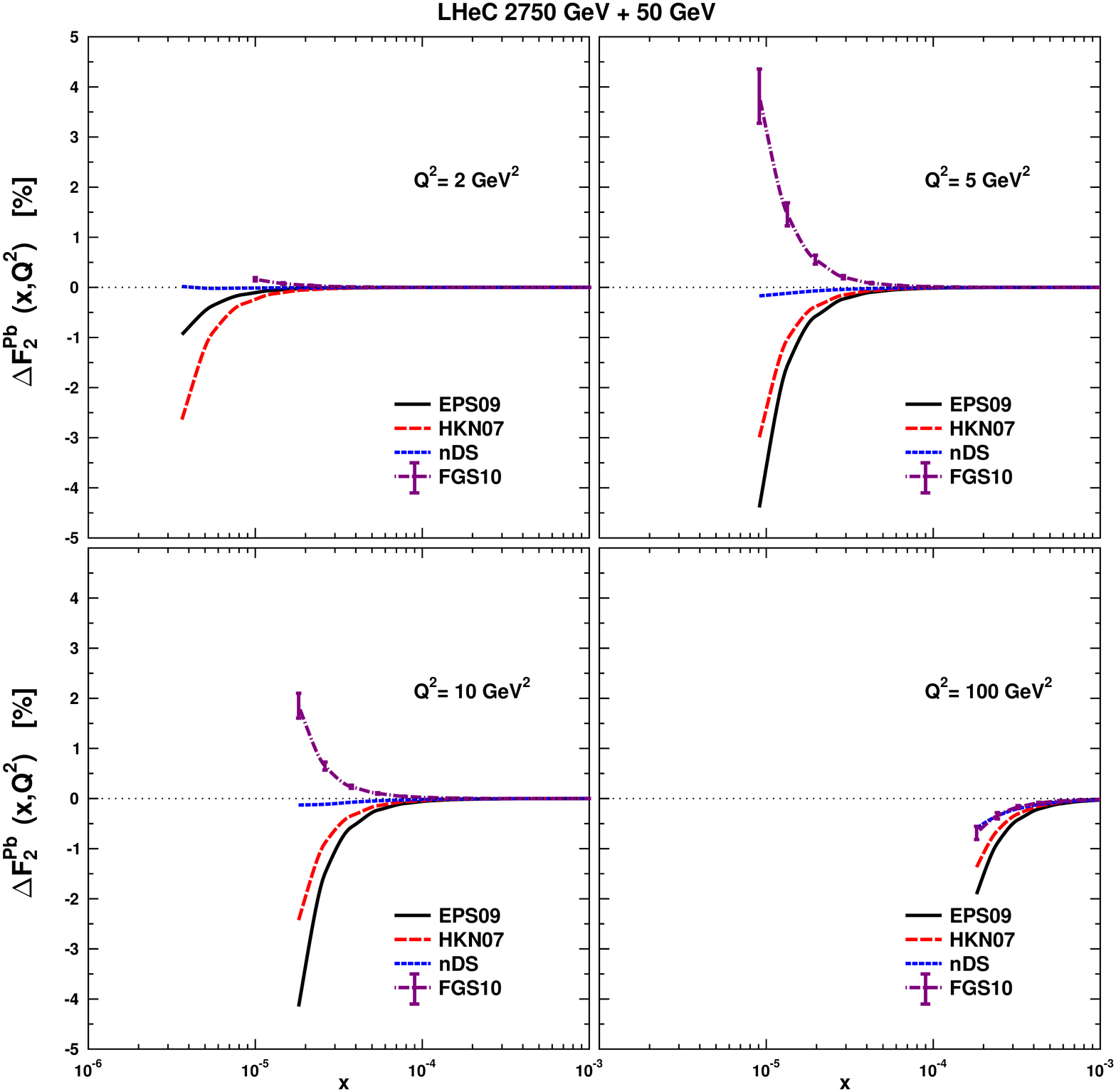}
\caption{Comparison of the $F_2$ extraction-uncertainty in EPS09, nDS, HKN07 and FGS10 for the LHeC kinematics. For the latter set, the error bars reflect the variation resulting from the two options provided in there.}
\label{fig:uncertainties2}
\end{center}
\end{figure}

\begin{figure}[!htb]
\begin{center}
\includegraphics[width=1.0\textwidth]{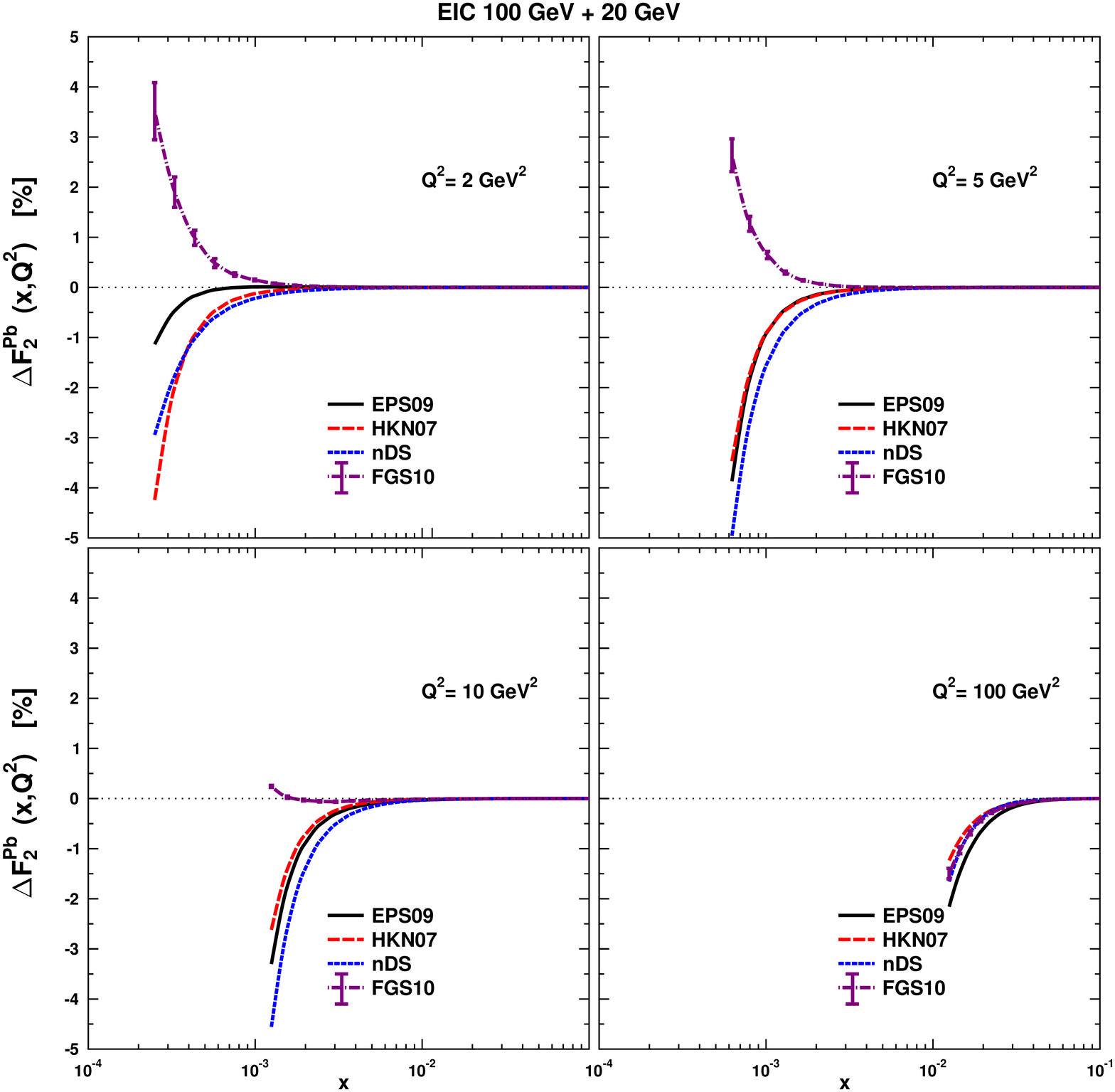}
\caption{Comparison of the $F_2$ extraction-uncertainty in EPS09, nDS, HKN07 and FGS10 for the EIC kinematics. For the latter set, the error bars reflect the variation resulting from the two options provided in there.}
\label{fig:uncertainties3}
\end{center}
\end{figure}

We show the results for Pb in Figs. \ref{fig:uncertainties}-\ref{fig:uncertainties3}. We consider two kinematical situations, a 100 GeV/nucleon proton or nucleus on a 20 GeV electron, and a
2750 GeV/nucleon
proton or nucleus on a 50 GeV electron -- which should roughly correspond to those collisions to 
be studied at the EIC \cite{eic} and at the LHeC \cite{lhec} respectively. In Fig. \ref{fig:uncertainties}, the uncertainty band corresponds to the one given by the application of the Hessian method to the nPDFs in the EPS09 parametrization \cite{Eskola:2009uj}, see that reference for details. 
In Fig. \ref{fig:uncertainties2} and \ref{fig:uncertainties3} we present the same quantity for the central values of the latest three NLO analyses of nuclear PDFs: EPS09  \cite{Eskola:2009uj}, nDS \cite{deFlorian:2003qf}, HKN07 \cite{Hirai:2007sx}. The trend of all these analyses is always similar with  the different magnitude of the effect reflecting the different relative amount of gluon and quark shadowing for the three cases. Also shown in these figures are the results from the FGS10 parametrization \cite{Guzey:2009jr}. As mentioned before, this set of nPDFs is built following a different procedure: the initial condition is obtained from the diffractive PDFs obtained in DIS with protons using the Gribov model of shadowing. Interestingly, this produces a correction with opposite sign to all other central values. 

In order to understand qualitatively the numerical behaviour seen in Figs.~\ref{fig:uncertainties}-\ref{fig:uncertainties3}, we make the approximation $F_L \big/x \simeq g(c x)$ and similarly $F_L^A\big/ x \simeq R^A_g(c x) g(c x)$, where $c \sim 2$ \cite{Altarelli:1978tq,Gluck:1978ky}, obtaining
\beq
\Delta F^A_2 (y = 1)\sim -\,\frac{R_g^A(cx)-R_{F_2}^A(x)}{R_{F_2}^A(x)} \,\frac{F_L(x)}{F_2(x)}\;,
\label{eq:deltaf2approx}
\eeq
where we have assumed that $F_2^A\gg F_L^A$ and the structure functions and nuclear ratios are to be evaluated at the virtuality corresponding to $y=1$.
Then, if the $Q^2$-evolution of the gluon ratio is slower (faster) than that of the sea, the result turns out to be positive (negative) even at initial low scales and increases (decreases) initially with the scale but finally,
at large $Q^2$, it vanishes as shadowing dies out logarithmically with $Q^2$. Examples of both behaviors can be seen in the figures.

This estimate also helps to understand the different behavior of FGS10 as compared with the other sets. According to Eq. (\ref{eq:deltaf2approx}), a gluon shadowing being much stronger than the corresponding one for quarks in the whole range of virtualities studied -- as it is the case for FGS10 -- would translate into a positive value of $\Delta F_2^A$. Notice that in a global fit, this quantity is also related with the logarithmic $Q^2$-slope of $R_{F_2}^A(x,Q^2)$ \cite{Eskola:2002us,Eskola:2008ca} 
which is constrained by existing data only for larger values of $x$ than the ones studied here.


On the other hand, the fact that for fixed $x$, $\Delta F_2^A$ decreases with increasing $Q^2$, is due to the kinematical factor $y^2/Y_+$ in Eqs. (\ref{relun}) and (\ref{rrr}), with $y=Q^2/(x s)$, $s$ being the Mandelstam variable of the lepton-hadron collision. Finally, the different behavior with energy of $|\Delta F_2^A|$  for fixed $Q^2$ and $y$, of the different sets (i.e. some of them show an increase, some of them a decrease), is due to specific features in the sets:  the different slopes in $x$ of the gluon and sea quark distributions, and their behavior at small $x$.

While obtaining the analogous results in the framework of saturation models (see Section \ref{intro}) requires a dedicated study out of the scope of this work, some qualitative estimates can be done. Using the dipole model (see the corresponding expressions in e.g. \cite{Armesto:2002ny}) and assuming scaling of the DIS cross sections with the variable $Q^2/Q_s^2$ with $Q_s^2$ some squared saturation momentum which increases with increasing nuclear size and with decreasing $x$ \cite{Stasto:2000er,Armesto:2004ud} -- as found in data and in non-linear QCD evolution --,  the ratio of the longitudinal to the transverse DIS cross sections, $\sigma_L/\sigma_T=(F_L/F_2)/(1-F_L/F_2)$, turns out to be generically a decreasing function of the scaling variable. Therefore, an increase in nuclear size (or a decrease in $x$) for fixed $Q^2$ would imply a shift of the scaling variable towards smaller values and thus an increase in the ratio\footnote{Using the results in \cite{Armesto:2002ny} where it was found that the nuclear effects on the longitudinal-to-transverse ratio are smaller than $\sim 10$ \% for Pb, the corresponding uncertainty $\Delta F^A_2 (y = 1)$ turns out to be negative and order 10 \%.}.

All in all, the uncertainties introduced by the nuclear effects are sizable, rising up to $\sim10$ \%, above all for small to moderate $Q^2$ and small $x$ - as expected. This stresses the need of either measuring the longitudinal structure functions for nuclei or providing experimental results for the full DIS cross section in future experimental programs on lepton-nucleus collisions \cite{eic,lhec}.

\section{Conclusions}
\label{conclu}

We have calculated nuclear ratios, defined in Eq.~(\ref{nsf}), with the most recent nPDFs for both nucleon and nuclei, shown the close correspondence between the nuclear effects on the glue and on $F_L$, and found significant nuclear size dependence of the ratio $F^A_L\big/ F_2^A$ at small $x$. We went on to demonstrate how this theoretical uncertainty will effect the experimental extraction of the nuclear structure function $F_2^A$. The resulting errors are largest, as high as $\sim 10\%$, in the most interesting kinematical region, namely at small $x$ and moderate $Q^2$ where future data can constrain the nuclear gluon distribution. This stresses the need for providing experimental data for nuclear DIS at future colliders in terms of reduced cross sections (total, charm and bottom), or, preferably, perform a collision energy scan to experimentally extract the nuclear longitudinal structure function.

\section*{Acknowledgements}

We thank M. Klein for useful discussions and V. Gon\c calves for comments. This work has been supported by Ministerio de Ciencia e Innovaci\'on of Spain under projects FPA2008-01177 and FPA2009-06867-E  and contracts Ram\'on y Cajal (NA and CAS), by Xunta de Galicia (Conseller\'{\i}a de Educaci\'on) and through grant PGIDIT07PXIB206\-126PR (NA and KT), grant INCITE08PXIB296116PR (CAS) and European Commission grant PERG02-GA-2007-224770 (CAS), and by the Spanish Consolider-Ingenio 2010 Programme CPAN (CSD2007-00042) (NA, HP and CAS).


\begin{thebibliography}{99}

\bibitem{Arneodo:1992wf}
  M.~Arneodo,
  Phys.\ Rept.\  {\bf 240} (1994) 301.

\bibitem{Geesaman:1995yd}
  D.~F.~Geesaman, K.~Saito and A.~W.~Thomas,
  Ann.\ Rev.\ Nucl.\ Part.\ Sci.\  {\bf 45} (1995) 337.
  
\bibitem{Armesto:2006ph}
  N.~Armesto,
  J.\ Phys.\ G {\bf 32} (2006) R367
  [arXiv:hep-ph/0604108].
  
\bibitem{Dokshitzer:1977sg}
  Y.~L.~Dokshitzer,
  Sov.\ Phys.\ JETP {\bf 46} (1977) 641
  [Zh.\ Eksp.\ Teor.\ Fiz.\  {\bf 73} (1977) 1216].

\bibitem{Gribov:1972ri}
  V.~N.~Gribov and L.~N.~Lipatov,
  Sov.\ J.\ Nucl.\ Phys.\  {\bf 15} (1972) 438
  [Yad.\ Fiz.\  {\bf 15} (1972) 781].

\bibitem{Altarelli:1977zs}
  G.~Altarelli and G.~Parisi,
  Nucl.\ Phys.\  B {\bf 126} (1977) 298.

\bibitem{Frankfurt:1990xz}
  L.~L.~Frankfurt, M.~I.~Strikman and S.~Liuti,
  Phys.\ Rev.\ Lett.\  {\bf 65} (1990) 1725.

\bibitem{Eskola:1992zb}
  K.~J.~Eskola,
  Nucl.\ Phys.\  B {\bf 400} (1993) 240.
  
\bibitem{Eskola:1998iy}
  K.~J.~Eskola, V.~J.~Kolhinen and P.~V.~Ruuskanen,
  Nucl.\ Phys.\  B {\bf 535} (1998) 351
  [arXiv:hep-ph/9802350].
  
\bibitem{Eskola:1998df}
  K.~J.~Eskola, V.~J.~Kolhinen and C.~A.~Salgado,
  Eur.\ Phys.\ J.\  C {\bf 9} (1999) 61
  [arXiv:hep-ph/9807297].

\bibitem{Collins:1989gx}
  J.~C.~Collins, D.~E.~Soper and G.~Sterman,
  Adv.\ Ser.\ Direct.\ High Energy Phys.\  {\bf 5} (1988) 1
  [arXiv:hep-ph/0409313].

\bibitem{Qiu:2003cg}
  J.~w.~Qiu,
  arXiv:hep-ph/0305161.

\bibitem{deFlorian:2003qf}
  D.~de Florian and R.~Sassot,
  Phys.\ Rev.\  D {\bf 69} (2004) 074028
  [arXiv:hep-ph/0311227].

\bibitem{Hirai:2007sx}
  M.~Hirai, S.~Kumano and T.~H.~Nagai,
  Phys.\ Rev.\  C {\bf 76} (2007) 065207
  [arXiv:0709.3038 [hep-ph]].

\bibitem{Eskola:2009uj}
  K.~J.~Eskola, H.~Paukkunen and C.~A.~Salgado,
  JHEP {\bf 0904} (2009) 065
  [arXiv:0902.4154 [hep-ph]].

\bibitem{Schienbein:2009kk}
  I.~Schienbein, J.~Y.~Yu, K.~Kovarik, C.~Keppel, J.~G.~Morfin, F.~Olness and J.~F.~Owens,
  Phys.\ Rev.\  D {\bf 80} (2009) 094004
  [arXiv:0907.2357 [hep-ph]].
 
\bibitem{Paukkunen:2010hb}
  H.~Paukkunen and C.~A.~Salgado,
  JHEP {\bf 1007} (2010) 032
  [arXiv:1004.3140 [hep-ph]].
  
\bibitem{Schienbein:2007fs}
  I.~Schienbein, J.~Y.~Yu, C.~Keppel, J.~G.~Morfin, F.~Olness and J.~F.~Owens,
  Phys.\ Rev.\  D {\bf 77} (2008) 054013
  [arXiv:0710.4897 [hep-ph]].

\bibitem{Guzey:2009jr}
  V.~Guzey and M.~Strikman,
  Phys.\ Lett.\  B {\bf 687} (2010) 167
  [arXiv:0908.1149 [hep-ph]].
  
\bibitem{Tywoniuk:2007xy}
  K.~Tywoniuk, I.~Arsene, L.~Bravina, A.~Kaidalov and E.~Zabrodin,
  Phys.\ Lett.\  B {\bf 657} (2007) 170
  [arXiv:0705.1596 [hep-ph]].

\bibitem{Armesto:2010kr}
  N.~Armesto, A.~B.~Kaidalov, C.~A.~Salgado and K.~Tywoniuk,
  Eur.\ Phys.\ J.\  C {\bf 68} (2010) 447
  [arXiv:1003.2947 [hep-ph]].
  
\bibitem{:2008tx}
  F.~D.~Aaron {\it et al.}  [H1 Collaboration],
  Phys.\ Lett.\  B {\bf 665} (2008) 139
  [arXiv:0805.2809 [hep-ex]].
  
\bibitem{Chekanov:2009na}
  S.~Chekanov {\it et al.}  [ZEUS Collaboration],
  Phys.\ Lett.\  B {\bf 682} (2009) 8
  [arXiv:0904.1092 [hep-ex]].
  
\bibitem{Thorne:2008aj}
  R.~S.~Thorne,
  arXiv:0808.1845 [hep-ph].
  
\bibitem{Frankfurt:2005mc}
  L.~Frankfurt, M.~Strikman and C.~Weiss,
  Ann.\ Rev.\ Nucl.\ Part.\ Sci.\  {\bf 55} (2005) 403
  [arXiv:hep-ph/0507286].
  
\bibitem{Ackerstaff:1999ac}
  K.~Ackerstaff {\it et al.}  [HERMES Collaboration],
  Phys.\ Lett.\  B {\bf 475} (2000) 386
  [Erratum-ibid.\  B {\bf 567} (2003) 339]
  [arXiv:hep-ex/9910071].
  
\bibitem{Cazaroto:2008qh}
  E.~R.~Cazaroto, F.~Carvalho, V.~P.~Goncalves and F.~S.~Navarra,
  Phys.\ Lett.\  B {\bf 669} (2008) 331
  [arXiv:0804.2507 [hep-ph]].
  
\bibitem{:2009wt}
  F.~D.~Aaron {\it et al.}  [H1 Collaboration and ZEUS Collaboration],
  JHEP {\bf 1001} (2010) 109
  [arXiv:0911.0884 [Unknown]].

\bibitem{Whitlow:1990gk}
  L.~W.~Whitlow, S.~Rock, A.~Bodek, E.~M.~Riordan and S.~Dasu,
  Phys.\ Lett.\  B {\bf 250} (1990) 193.
  
  \bibitem{eic}
The Electron Ion Collider Working Group Collaboration, C.~Aidala {\it
  et al.}, {\it {A High Luminosity, High Energy Electron Ion Collider}}, 
  {\tt http://web.mit.edu/eicc/}.

\bibitem{lhec} M.~Klein {\it et al.}, {\it {Prospects for a Large
      Hadron Electron Collider (LHeC) at the LHC}}, EPAC'08, 11th
  European Particle Accelerator Conference, 23- 27 June 2008, Genoa,
  Italy; {\tt http://www.lhec.org.uk/}.
  
\bibitem{Albacete:2009fh}
  J.~L.~Albacete, N.~Armesto, J.~G.~Milhano and C.~A.~Salgado,
  Phys.\ Rev.\  D {\bf 80} (2009) 034031
  [arXiv:0902.1112 [hep-ph]].

\bibitem{Armesto:2002ny}
  N.~Armesto,
  Eur.\ Phys.\ J.\  C {\bf 26} (2002) 35
  [arXiv:hep-ph/0206017].
  
\bibitem{Kugeratski:2005gx}
  M.~S.~Kugeratski, V.~P.~Goncalves and F.~S.~Navarra,
  Eur.\ Phys.\ J.\  C {\bf 46} (2006) 465
  [arXiv:hep-ph/0508255].
  
\bibitem{Cazaroto:2008iy}
  E.~R.~Cazaroto, F.~Carvalho, V.~P.~Goncalves and F.~S.~Navarra,
  Phys.\ Lett.\  B {\bf 671} (2009) 233
  [arXiv:0805.1255 [hep-ph]].

\bibitem{Stump:2003yu}
  D.~Stump, J.~Huston, J.~Pumplin, W.~K.~Tung, H.~L.~Lai, S.~Kuhlmann and J.~F.~Owens,
  JHEP {\bf 0310} (2003) 046
  [arXiv:hep-ph/0303013].
  
\bibitem{Brock:1993sz}
  R.~Brock {\it et al.}  [CTEQ Collaboration],
  Rev.\ Mod.\ Phys.\  {\bf 67} (1995) 157.
  
\bibitem{Altarelli:1978tq}
  G.~Altarelli and G.~Martinelli,
  Phys.\ Lett.\  B {\bf 76} (1978) 89.

\bibitem{Gluck:1978ky}
  M.~Gluck and E.~Reya,
  Nucl.\ Phys.\  B {\bf 145} (1978) 24.

\bibitem{Eskola:2002us}
  K.~J.~Eskola, H.~Honkanen, V.~J.~Kolhinen and C.~A.~Salgado,
  Phys.\ Lett.\  B {\bf 532} (2002) 222
  [arXiv:hep-ph/0201256].

\bibitem{Eskola:2008ca}
  K.~J.~Eskola, H.~Paukkunen and C.~A.~Salgado,
  JHEP {\bf 0807} (2008) 102
  [arXiv:0802.0139 [hep-ph]].
  
\bibitem{Stasto:2000er}
  A.~M.~Stasto, K.~J.~Golec-Biernat and J.~Kwiecinski,
  Phys.\ Rev.\ Lett.\  {\bf 86} (2001) 596
  [arXiv:hep-ph/0007192].

\bibitem{Armesto:2004ud}
  N.~Armesto, C.~A.~Salgado and U.~A.~Wiedemann,
  Phys.\ Rev.\ Lett.\  {\bf 94} (2005) 022002
  [arXiv:hep-ph/0407018].

\end{thebibliography}
\end{document}